\documentstyle[12pt,epsfig,equation]{article}
\begin{document}
\newcommand{\be}{\begin{equation}}
\newcommand{\ben}{\begin{subequations}}
\newcommand{\een}{\end{subequations}}
\newcommand{\beq}{\begin{eqalignno}}
\newcommand{\eeq}{\end{eqalignno}}
\newcommand{\ee}{\end{equation}}
\newcommand{\nav}{\langle n \rangle}
\renewcommand{\thefootnote}{\fnsymbol{footnote}}
\pagestyle{empty}
\begin{flushright}
APCTP 97--03 \\
March 1997\\
\end{flushright}
\vspace*{1.5cm}

\begin{center}
{\Large \bf Are the H1 and ZEUS ``High $Q^2$ Anomalies'' Consistent with
Each Other?} \\
\vspace*{5mm}

Manuel Drees \\
{\it APCTP, 207--43 Cheongryangri--dong, Tongdaemun-gu, Seoul 130--012, Korea}
\end{center}

\begin{abstract}
Both ZEUS and H1 have recently reported an excess of events at high $Q^2$ and
high Bjorken--$x$. However, the $x$ distributions differ 
considerably; moreover, H1 sees more events with lower luminosity. Taken
separately, the $x$ distributions and the number of observed events are 
consistent between the two experiments at the few percent level. However, when
combined it becomes clear that the results of H1 and ZEUS are as
(in)compatible with each other as each is with the Standard Model.

\end{abstract}
\clearpage
\setcounter{page}{1}
\pagestyle{plain}

Very recently both experiments at the $ep$ collider HERA, H1 \cite{1}
and ZEUS \cite{2}, announced an excess of events that look like
deep--inelastic scattering (DIS) events with high squared 4--momentum
$Q^2$ and high Bjorken$-x$. Both experiments estimated the probability
of getting such an excess from a fluctuation within the Standard Model
(SM) to be somewhat less than 1\%. Naively the combination of both
experiments therefore excludes the SM at the better than 99.99\%
confidence level. This is causing a flurry of papers [3--9] that
attempt to explain the effect in terms of some ``new physics''.

However, combining the results from the two experiments only makes sense
if they are consistent with each other. In this note I argue that the level
of consistency of the two experiments with each other is no better than
the compatibility of each experiment by itself with the SM. The two
results can therefore not be claimed to mutually reinforce each other.

In order to arrive at this conclusion, one has to examine two aspects of the
data sets: The distributions in Bjorken-$x$, and the number of excess
events seen by H1 and ZEUS. Let me begin with the $x$ distributions.

The fact that these distributions look different has been noted before 
\cite{5,6,9}; H1 seems to see a rather well--defined peak at 
$M_e = \sqrt{x_e s} \simeq 200$ GeV, whereas the ZEUS events are broadly
distributed around 220 GeV. Here $\sqrt{s} =300$ GeV is the $ep$
centre--of--mass energy, and the subscript ``$e$'' indicates that the
momentum of the outgoing positron has been used to determine
$x$. It has been argued \cite{5,6} that comparing the two distributions
might be misleading, since ZEUS uses a different method (the ``double
angle'' method) to reconstruct $x$. However, this is not entirely correct.
While ZEUS {\em prefers} the double angle method, since it gives more
precise results given the performance of their detector, they do also
give results for $x_e$. Table 1a,b lists the events seen by H1 and
ZEUS, respectively, in the H1 search window, defined by $M_e \geq 180$
GeV, $y_e > 0.4$. ($y$ is the scaled positron energy loss in the proton
rest frame.)

\vspace{1cm}
\noindent

\begin{center}
a) H1 events \\
\begin{tabular}{|c|c|c|c|c|c|c|c|}
\hline
\# & 1 & 2 & 3 & 4 & 5 & 6 & 7 \\
\hline
$M_e$ & $196 \pm 5$ & $208 \pm 4$ & $188 \pm 12$ & $198 \pm 2$ & $211 \pm 4$ & 
$192 \pm 6$ & $200 \pm 2$ \\
\hline
\end{tabular}
\end{center}

\vspace{.5cm}
\noindent
\begin{center}
b) ZEUS events\\
\begin{tabular}{|c|c|c|c|c|}
\hline
\# & 1 & 2 & 3 & 4 \\
\hline
$M_e$ & $218 \pm 10$ & $220 \pm 10$ & $234 \pm 12$ & $200 \pm 14$ \\
\hline
\end{tabular}
\end{center}

\vspace{.5cm}
\noindent
{\bf Table 1:} The $M_e$ values in GeV of the events found by H1 (a) and 
ZEUS (b) in the region $M_e \geq 180$ GeV, $y_e \geq 0.4$.

\vspace{1cm}

As stated earlier, the ZEUS events seem to lie at higher values of $M_e$.
Specifically, the averages of the two distributions are:
\ben \label{e1} \beq
\overline{M_e}({\rm H1}) &= ( 200.3 \pm 1.2) \ {\rm GeV}; \label{e1a} \\
\overline{M_e}({\rm ZEUS}) &= (219.3 \pm 5.5) \ {\rm GeV}. \label{e1b}
\eeq \een
These two values seem to differ by 3.4 standard deviations! However, so
far I have only included the errors that are uncorrelated from event to
event. Both H1 and ZEUS quote an overall energy scale uncertainty of
3\% for their electromagnetic calorimeters. Call $r$ the energy scale
factor. Then \cite{1}
\be \label{enew}
\left. \frac {1} {M_e} \frac {d M_e} {d r} \right|_{r=1} = 1 +
\frac {1-2y_e} {2y_e}
\ee
The second term can be neglected for $y_e$ near 0.5, where most of the
events are; neglecting it for the few events at high $y_e$ is conservative,
since it will overestimate the impact of the energy scale uncertainty on
the determination of $\overline{M_e}$. With this assumption, a 3\% energy scale
uncertainty simply gives a 3\% uncertainty in $\overline{M_e}$. Adding this 
systematic error in quadrature to the statistical errors in eqs.(\ref{e1}),
and assuming that the H1 and ZEUS energy scale factors are independent of
each other, one finds:
\be \label{e2}
\overline{M_e}({\rm ZEUS}) - \overline{M_e}({\rm H1}) = (19.0 \pm 10.5)
\ {\rm GeV},
\ee
which amounts to a difference of 1.8 ``standard deviations''. Naively this
means that the two distributions are compatible with each other at only the
8\% level. However, since the error in eq.(\ref{e2}) is mostly due to 
systematic effects (the energy scale uncertainties of both experiments), no
straightforward statistical interpretation can be given.

One can overcome the scale uncertainty by determining the variance 
$\sigma_{M_e}$ of the distributions (or, equivalently, their second moments).
From table 1 one finds:
\ben \label{e3} \beq
\sigma_{M_e}({\rm H1}) &= (3.3 \pm 0.6) \ {\rm GeV}; \label{e3a} \\
\sigma_{M_e}({\rm ZEUS}) &= (9.1 \pm 4.1) \ {\rm GeV}, \label{e3b}
\eeq \een
which gives
\be \label{e4}
\sigma_{M_e}({\rm ZEUS}) - \sigma_{M_e}({\rm H1}) = (5.8 \pm 4.2) \ 
{\rm GeV}.
\ee
This only amounts to a 1.4 standard deviation ``discrepancy'', due to the
larger errors of ZEUS' $M_e$ values, and the rather small number of events
seen by ZEUS, which does not really suffice to determine the shape of the
$M_e$ distribution.

This brings me to the second part of my argument, the discrepancy in the 
number of events seen by the two experiments. In the search region
$M_e \geq 180$ GeV, $y_e \geq 0.4$ defined by H1, they find seven events, but
ZEUS only finds four.\footnote{ZEUS lists a fifth excess event, at 
$y_e = 0.32$; however, ZEUS does not seem to have a well--defined search 
region. This fifth event has $M_e = (225 \pm 20)$ GeV; its inclusion would
therefore make the discrepancy between the two $M_e$ distributions slightly
worse.} This is in spite of the fact that ZEUS analyzed 30\% more data
($\int {\cal L} dt = 20.1$ pb$^{-1}$, vs. 14.2 pb$^{-1}$ for H1), {\em and}
used looser cuts that result in a higher efficiency. Comparing the number
of events with $Q^2 \geq Q^2_{\rm min}$ expected within the SM given by
the two experiments (table 2 of \cite{1} and table 6 of \cite{2}), one
concludes
\be \label{e5}
c \equiv \frac { \nav_{\rm H1} } {\nav_{\rm ZEUS} } = 0.55 \ {\rm to} \  
0.60,
\ee
where $\nav = \sigma \epsilon \int {\cal L} dt$ is the expected number
of events for given cross section $\sigma$, luminosity $\int {\cal L} dt$
and efficiency $\epsilon$. Here I am assuming that the efficiency for
putative non--SM events in the samples is the same as for SM DIS events
at high $Q^2$. The authors of ref.\cite{5} have checked that this is 
indeed the case for events due to the $s-$channel production of
leptoquarks (or squarks with appropriate R--parity breaking couplings,
which amounts to the same thing in this context), which is the so far most
popular interpretation of the excess [4--8]. Note also that the ratio
$c$ for DIS events is \cite{1,2} essentially independent of $Q^2_{\rm min}$ 
for $Q^2_{\rm min} \geq 5,000$ GeV$^2$.

From eq.(\ref{e5}) one concludes that if $\nav_{\rm H1} = n_{\rm H1} = 7$,
then $\nav_{\rm ZEUS}=11$ to 13, compared to the actual $n_{\rm ZEUS} = 4$.
This shows that the hypothesis $\nav_{\rm H1} = 7$ is excluded by the
ZEUS data at the 99\% (99.5\%) c.l. for $c=0.6 \ (0.55)$. Simply taking
\cite{4} the number of excess events observed by H1 as a measure for any
``new physics'' cross section is therefore misleading.

However, this does not directly tell us how (in)compatible the number of
events seen by the two groups are. In order to investigate this question
one has to let $\nav_{\rm H1}$ or, equivalently, $\nav_{\rm ZEUS}$
float, and compute the combined probability
\be \label{e6}
P \left( \nav_{\rm ZEUS},c \right) = \frac { p(n_{\rm H1} \geq 7) \cdot
p(n_{\rm ZEUS} \leq 4) } {p_{\rm max}},
\ee
where
\ben \label{e7} \beq
p(n_{\rm H1} \geq k) &= 1 - \sum_{n=0}^{k-1} \frac { \left( 
c \nav_{\rm ZEUS} \right)^n} {n!} e^{- c \nav_{\rm ZEUS}}; \label{e7a} \\
p(n_{\rm ZEUS} \leq k) &= \sum_{n=0}^{k} \frac { \left( 
\nav_{\rm ZEUS} \right)^n} {n!} e^{- \nav_{\rm ZEUS}}; \label{e7b} \\
p_{\rm max} &= \max_j \left\{ p(n_{\rm ZEUS} \leq j) \cdot p(n_{\rm H1} 
\geq [ c \cdot j ]) \right\} . \label{e7c}
\eeq \een
The interpretation of eqs.(\ref{e7}a,b) is straightforward. The
denominator $p_{\rm max}$ in eq.(\ref{e6}), given in eq.(\ref{e7c}),
is the maximal value of the numerator for {\em any} combination of
$n_{\rm ZEUS}$ and $n_{\rm H1}$, assuming only that $n_{\rm H1} \geq
[c \cdot n_{\rm ZEUS}]$\footnote{In the opposite case, i.e.  if ZEUS
had seen ``too many'' events compared to H1, the inequalities in the
numerator in eq.(\ref{e6}) would have to be reversed.} where $[ x ]$
refers to taking the integer part of $x$ and $c$ is defined in
eq.(\ref{e5}). Note that this maximum is obviously always reached if
the number of H1 events is as small as allowed by this constraint;
hence only the number of ZEUS events $j$ needs to be scanned in
eq.(\ref{e7c}). This normalization ensures that $P$ reaches unity for
some range of $\nav_{\rm ZEUS}$ in the simple test case where both
experiments see the same number of events and $c=1$, in agreement with
an intuitive definition of the compatibility of the two experiments.
Numerically, $p_{\rm max} \simeq 0.35$ to 0.4 for the case at hand.

The dependence of $P$ on $\nav_{\rm ZEUS}$ is shown in Fig.~1. One concludes
that the number of excess events seen by the two groups are compatible with
each other only at the 5.2\% (3.7\%) level for $c=0.60$ (0.55). This
conclusion is completely independent of the $M_e$ distributions discussed
above. If the error in eq.(\ref{e2}) could be interpreted as simple
statistical uncertainty, one would conclude that the overall probability
that the two experiments are consistent with each other is less than
0.5\%. This number is slightly smaller than the probability that either one 
of the experiments is compatible with the SM. The predominance of the
systematic error in eq.(\ref{e2}) makes it difficult to give such a precise
estimate of the level of (in)compatibility of the results of the two
experiments; however, it seems clear that it cannot be much above the
1\% level.

This means that {\em any} interpretation of the existing data sets
that assumes them to be statistically independent {\em has} to assume that
at least one fluctuation occurred which had an a priori probability of no
more than 1\%.\footnote{More exactly, one needs two independent
fluctuations, in the event numbers and $M_e$ distributions, with combined
probability of 1\% or less.} Of course, it still remains true that an
interpretation within the SM would need {\em two} such fluctuations.
However, since one fluctuation of this kind evidently has to have happened
anyway,\footnote{The only way around this conclusion seems to require the
assumption that the H1 and ZEUS data sets are somehow {\em not}
statistically independent. This, however, quickly takes one into the realm
of conspiracy theories.} one can at present only conclude that the
recent HERA data disfavor the SM at the 99\% level, as opposed to the
99.99\% level. Of course, an ``exclusion'' of the SM at the 99\% c.l. 
is not very impressive. For example, both the L3 ``$l^+ l^- \gamma \gamma$''
events \cite{l3} and the ALEPH ``$\tau^+ \tau^- V$'' events \cite{aleph}
``excluded'' the SM at significantly higher confidence level, and nevertheless
later proved to be spurious (most likely due to fluctuations). I would
therefore not be surprised at all if the SM weathered this most recent
onslaught as well.

\noindent
\setcounter{figure}{0}
\begin{figure}
\vspace*{2.6cm}
\centerline{\epsfig{file=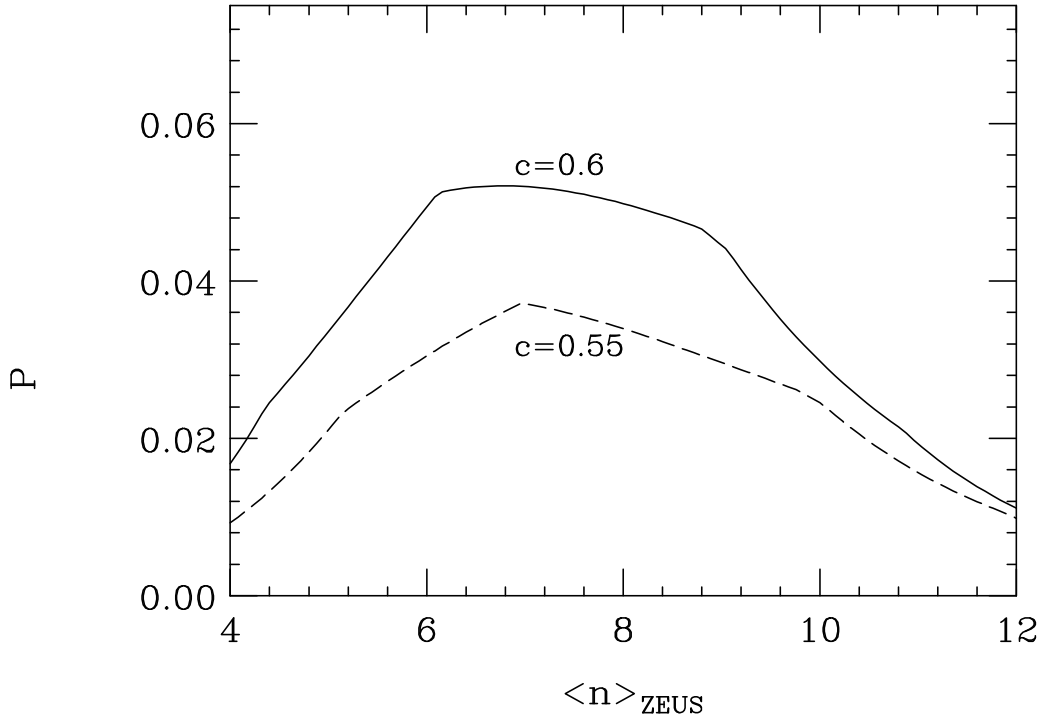,height=10cm}}
\caption
{The probability $P$ defined in eq.(6) is shown as a function of the
{\em expected} number of events that should have been seen by the ZEUS
experiment for given cross section. $c = \langle n \rangle_{\rm H1} / 
\langle n \rangle_{\rm ZEUS}$, see eq.(5).}
\end{figure}

 \end{document}